\documentclass[12pt,a4]{article}
\usepackage{amsmath}
\usepackage{amssymb}
\usepackage{amsfonts}
\usepackage{array}
\usepackage{graphicx}% use this package if an eps figure is included.
\usepackage{mathrsfs}
\usepackage{multirow}
\usepackage{siunitx}
\usepackage{hyperref}
\usepackage{xcolor}
\newcommand{\bea}{\begin{eqnarray*}}
\newcommand{\eea}{\end{eqnarray*}}
\newcommand{\beao}{\begin{eqnarray}}
\newcommand{\eeao}{\end{eqnarray}}

\hypersetup{ 
colorlinks=true,
filecolor=magenta,
urlcolor=cyan,
    pdfauthor={Gernot Deinzer},
    pdftitle={Cost for research – how cost data of research can be included in open metadata to be reused and evaluated},
    pdfsubject={Information Science},
    pdfkeywords={cost data, cost transparency, OAI-PMH, xml schema, sustainable publishing}
}

\begin{document}
\title{Cost for research – how cost data of research can be included in open metadata to be reused and evaluated}

\author{
	Bartlewski Julia\href{https://orcid.org/0000-0001-5959-4999}{\includegraphics[scale=.8]{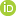}} \textsuperscript{1}
,  
Christoph Broschinski\href{https://orcid.org/0000-0003-1972-7587}{\includegraphics[scale=.8]{orcid_16x16.png}} \textsuperscript{1}
, 
Gernot Deinzer %\orcidlink{0000-0000-0000-0000}
 \href{https://orcid.org/0000-0002-7462-3847}{\includegraphics[scale=.8]{orcid_16x16.png}} \textsuperscript{2},
\\
Cornelia Lang \href{https://orcid.org/0000-0001-7046-344}{\includegraphics[scale=.8]{orcid_16x16.png}} \textsuperscript{2},
Dirk Pieper\href{https://orcid.org/0000-0001-5959-4999}{\includegraphics[scale=.8]{orcid_16x16.png}} \textsuperscript{1},
Bianca Schweighofer\href{https://orcid.org/0000-0002-9416-9311}{\includegraphics[scale=.8]{orcid_16x16.png}} \textsuperscript{2},
Colin Sippl\href{https://orcid.org/0000-0002-8503-1740}{\includegraphics[scale=.8]{orcid_16x16.png}} \textsuperscript{2},
\\
Lisa-Marie Stein\href{https://orcid.org/0000-0001-7905-0462}{\includegraphics[scale=.8]{orcid_16x16.png}} \textsuperscript{3},
Alexander Wagner\href{https://orcid.org/0000-0001-9846-5516}{\includegraphics[scale=.8]{orcid_16x16.png}} \textsuperscript{3},
Silke Weisheit\href{https://orcid.org/0000-0002-2609-4274}{\includegraphics[scale=.8]{orcid_16x16.png}} \textsuperscript{2}
\\
\small  \textsuperscript{1}University of Bielefeld, Germany, \textsuperscript{2}University of Regensburg, Germany, \textsuperscript{3}DESY, Hamburg, Germany
}
\maketitle

\begin{abstract}

The openCost project aims to enhance transparency in research funding by making publication-related costs publicly accessible, following FAIR principles. It introduces a metadata schema for cost data, allowing aggregation and analysis across institutions. The project promotes open access and cost-efficient models, benefiting academic institutions, funders, and policymakers.

\end{abstract}

Licence: \href{https://creativecommons.org/licenses/by/4.0/}{\includegraphics[scale=.8]{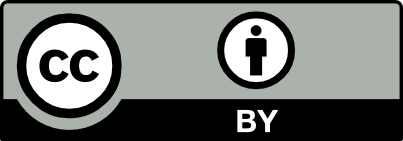}}

\section*{Purpose}

In the contemporary context of research, financial expenditures are escalating at an alarming rate. 
This encompasses not only direct expenses associated with the research process, such as staff, hardware, and equipment, but also the dissemination and archiving of results. Usually, these costs remain inaccessible to the public, a state of affairs that is deemed to be highly unsatisfactory. Consequently, we assert that costs must be rendered publicly accessible in a manner analogous to research data, in accordance with the FAIR principles \cite{1}. The openCost \cite{2} project exemplifies the feasibility of this objective by illustrating the publication of costs associated with the publication of an article. 

The utilisation of standard interfaces facilitates the aggregation of data, enabling its reuse for analysis of costs across institutions and subjects. The overarching objective is to achieve transparency in financial flows from academic institutions to various service providers, including commercial and non-commercial repository operators, long-term archiving initiatives, and infrastructure providers.

\section*{Methods}

The use of metadata in the form of well-known formats such as Dublin Core \cite{3} or DataCite \cite{4} Metadata Schema is commonplace in the documentation of journal articles, research data and all manner of scientific publications. Metadata employed within these frameworks can be categorised into distinct types:
\begin{itemize}
	\item bibliographic metadata
	\item legal metadata
	\item technical metadata 
	\item administrative metadata
\end{itemize}
In collaboration with international experts, openCost has developed a metadata schema to model costs associated with publications, research data, software, etc.

Taking the complex landscape of scientific publishing into account, we identified two distinct cases of cost association: fees directly associated with the published item and costs incurred by means of larger contracts that cover several publications (e.g. so called transformative agreements and memberships).

In the first case we propose to store cost information directly with the item as integral part of the published items metadata. Being agnostic to the actual item at hand this procedure is also  applicable to research data and software.
As there is no obvious way to address the total costs of articles published within transformative agreements, a new entity called contract is introduced. This entity holds all costs associated for items subject to the agreement. Articles will then be linked to the contract entity to allow for a transparent item based cost calculation. A  metadata set for the contract entity is proposed, with all the relevant information stored therein. The linking  allows for various evaluations of the cost efficiency of a given contract.
Being modeled in a general and again publication-agnostic way, the contract entity can be applied to diamond business models, consortial memberships and infrastructures such as the Directory of Open Access (DOAJ) \cite{5}, even in the absence of publications connected with it.

It is proposed that cost-relevant data has to be included in the metadata of institutional repositories, made publicly available without restrictions (CC-0), and provided in machine-readable form. To this end, an XML-representation as well as a JSON-representation of the new schema have been formulated. Institutional repositories are the optimal choice for cost data, as by their very nature they are able to store metadata in a standardised form and provide output interfaces such as the OAI-PMH \cite{6} interface. A harvesting interface like OAI-PMH seamlessly handles data updates and can be used efficiently to transfer even large sets of data regularly. Therefore it is ideally suitable for constant monitoring within an institution as well as beyond, which is desirable e.g. for national contracts. The repository is designed to hold all publications made available by researchers. It should be noted that in instances where open versions are not yet available, this is still possible, if there is an entry in the repository (maybe an access restricted version and a "request a copy" button can be utilised to retrieve a private copy). Hence cost information like colour, cover, or page charges, for closed access articles can be included in repositories as well. Additionally, we strongly advocate that academic institutions store copies of diamond and gold OA articles within their respective repositories to get an overall information of publication and related business models. 

By employing the OAI-PMH interface, the data is provided reusable by machines and can be harvested by service providers such as OpenAPC \cite{7}. Subsequent evaluation of the data across various institutions is also facilitated.

\section*{Results}
The proposed metadata scheme has the capacity to describes all cases of publication cost payments: an individual fee for a single publication (which must not be related to OA) and the paying for a group of articles e.g. within a transformative agreement.

The full documentation of the metadata schema can be found at github \cite{8}.

The proposed XML schema has been implemented in the repositories of the University of Regensburg (software EPrints \cite{9}), the University of Bielefeld (based on LibreCat \cite{10}) and the repositories of the JOIN2-consortia \cite{11}, such as at DESY. Other institutions, like the University of Ulm, are planning to integrate the schema in their repository based on Dspace 7 \cite{12}.

The metadata schema is already in use for harvesting data from DESY by OpenAPC with the goal to replace the  previously required submission of data via email or the upload to a GitHub repository. This approach eliminates the need for manual intervention from DESY and reduces the workload for OpenAPC. It enables continuous harvesting, ensuring the accuracy and currency of data in the OpenAPC database. Prior to this development, delays of up to a year were common.

The scalability of the openCost approach has been demonstrated by automatic data delivery from Forschungzentrum Jülich (FZJ) to OpenAPC. The corpus transferred holds data from over 100 academic institutions throughout Germany who are obliged to report publication cost expenditures to FZJ within the framework of the funding programme "Open Access Publication Funding" \cite{13} by the Deutsche Forschungsgemeinschaft (DFG).

The data of openAPC are reused by the Electronic Journal Library (Elektronische Zeitschriftenbibliothek EZB) \cite{14} to display global and local averages of publication costs for journals.

Further initiatives such as the "Transparency to Sustain Open Science Infrastructure" \cite{15} by the Université Grenoble Alpes and "Reasonable costs for public access" \cite{16} by the Invest in Open Infrastructures (IOI) show a strong interest in our project. Therefore, we believe in a global potential of our metadata schema. 

\section*{Value}

The incorporation of metadata related to costs into research processes aligns with the principles of Open Science, as outlined by the FAIR guidelines, by enhancing transparency. This approach not only addresses the absence of missing costs in the context of Open Science but also demonstrates significant benefits for various stakeholders. Notably, academic institutions profit from compliance with legal guidelines, such as open budget, and from fulfilling requirements research funders, like the DFG. Furthermore, this approach facilitates the comparison of specific costs with those of other institutions, offering a comprehensive view of expenditures. Funder organisations gain a more comprehensive understanding of their spending on publications, while the political sector benefits from enhanced transparency in research processes. The application of our approach, encompassing diamond and commercial publishing, was designed to be open for all scientific outputs and holds the potential to be adopted by the majority of academic institutions. This would lead to proven cost savings in diamond open access, thereby facilitating the transition of the publication system to diamond business models.

%\section*{References}

\end{document}